\documentclass[a4paper, times, 10pt, twocolumn, twoside,top=3.5cm,bottom=3.5cm,left=2.5cm,right=2cm]{article}

\usepackage{ISARC}
\usepackage{graphicx,subfigure}
\usepackage{color}
\usepackage{CJK}
\usepackage{framed}
\usepackage{ulem}
\usepackage{lscape}
\usepackage{hologo}

\begin{document}

\definecolor{shadecolor}{rgb}{1,1,0}

\linespread{0.5}

\title{Radiative Heat Transfer \\in 
Free-Standing Silicon Nitride Membranes}

\author{Chang Zhang, Mathieu Giroux, Thea Abdul Nour, Raphael St-Gelais*}

\affiliation{
Department of Mechanical Engineering, University of Ottawa, Ottawa, ON, Canada\\
$^*$raphael.stgelais@uottawa.ca
}

\maketitle 
\thispagestyle{fancy} \pagestyle{fancy}

\begin{abstract}
Free-standing silicon nitride (SiN) mechanical resonators are of central interests in applications such as temperature and mass sensing, and for fundamental optomechanical research. Understanding thermal coupling between a membrane resonator and its environment is required for predicting thermal noise, frequency noise, as well as sensors responses to temperature changes. In this work, we provide closed-form derivations of intrinsic thermal coupling quantities in free-standing thin films--namely total thermal conductance with the surroundings, thermal response time, and the relative contribution of thermal radiation. Our model is valid for any free-standing thin film anchored on all sides, although we particularly emphasize the specific case of SiN for which spectral emissivity is thoroughly investigated as a function of thickness and temperature. We find that radiative heat exchanges can play a non-negligible role, and even dominate thermal coupling for membranes of sizes commonly employed in optomechanics experiments. We experimentally confirm the validity of our model by measuring radiative thermal coupling between a SiN mechanical resonator and a ceramic heater in high vacuum. 
\end{abstract}

\section{Introduction}
\label{sec:Introduction}
Thin-film silicon nitride (SiN) membranes are heavily used as mechanical and optical resonators in both fundamental opto-mechanical studies \cite{Wilson2009,Regal2011,Chakram2013} and many state-of-the-art sensing technologies, including ultra-sensitive mass sensing \cite{Hanay2012}, gas detection \cite{JJ_Greffet}, nanoparticulate mass detection \cite{Larsen2013}, thermal radiation sensing \cite{Zhang2013,8956551} and pressure sensing \cite{Zhu2013}. Some of these works involve cooling \cite{Wilson2009,Chakram2013}, while in other cases, the temperature dependence of membrane stress is used as a sensing mechanism \cite{8956551,Zhang2013}, or an active technique for controlling resonator frequency \cite{St-Gelais2019}.

Despite, the strong influence of temperature in SiN membranes, there is still no closed-form expression describing thermal coupling with their environment, as well as their characteristic thermal response time ($\tau$). As such, our goal is to provide expression for $\tau$; for the heat conduction between the membrane and the environment ($G$, in $\mathrm{W/K}$); and for the fraction ($x_{rad}$) of this conduction that occurs via thermal radiation. We provide these expressions for the specific case SiN films of square and circular shape in vacuum (i.e., in the absence of convection heat transfer).

Analytical expression for thermal coupling of SiN membranes are notably needed for understanding their fundamental noise mechanisms. Values of $G$ and $\tau$ are central in the calculation of noise processes such as temperature fluctuation noise and temperature-induced frequency fluctuation \cite{Sansa2016,Fong2012,Cleland2002}.

Likewise, in the context of nanomechanical radiation sensors \cite{Zhang2013}, understanding the ratio of heat transfer occurring via radiation ($x_{rad}$) is key for determining ultimate sensor detectivity and noise equivalent power \cite{Rogalski2003}. The fundamental performance limit of a thermal-based radiation sensor is notably reached when thermal coupling between the sensor and the surroundings is dominated by radiation (i.e., $x_{rad}\approx1$) \cite{Zhang2013,St-Gelais2019,Blaikie2019}. 

Efforts have been devoted to investigating the thermal properties \cite{Sikora2012,Zhang1995,Tien2012} of SiN membranes such as thermal conductivity, heat capacity, diffusivity and thermal expansion coefficient. Emissivity of metal coating on SiN membranes have also been investigated in the context of electron microscopy \cite{VanZwol2015}. Meanwhile, others have provided finite element simulation \cite{Revaz2003} of temperature profiles SiN membranes. Nevertheless, to the best of our knowledge, there is still no rigorous closed-form analytical expression describing heat transfer in such membranes. 

\section{Heat transfer in free-standing thin films}
\begin{figure}[!htb]
    \centering
    \includegraphics[width=0.46\textwidth]{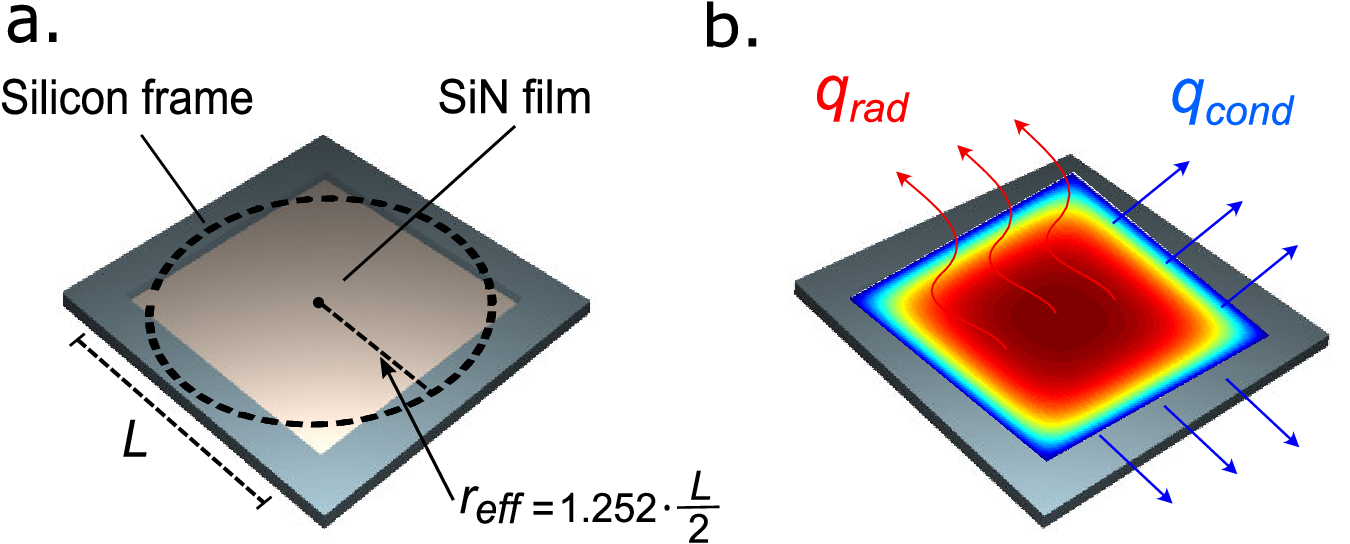}
    \caption{\label{fig:one} (a) Square-shape SiN membrane and silicon frame. (b) Surface temperature profile of the SiN film with uniform internal generation. $q_{rad}$ and $q_{cond}$ denote the heat flux transferred via radiation and conduction, respectively.}
 
\end{figure}

In order to calculate thermal coupling between a freestanding film and its environment, we consider volumetric heat generation ($\dot q,\textrm{in} \mathrm{W/m^3}$) occurring uniformly within the membrane. From a heat transfer standpoint, such internal generation is mathematically analogous to uniform absorption of radiation from an external radiation source (e.g., in the context of radiation sensors). As a result of this internal generation, membrane temperature ($T_{m}$) rises relative to ambient ($T_{\infty}$), and heat leaves the membrane by conduction to the silicon frame ($q_{cond}$, in $\text{W}$), and re-emission of radiation to the environment ($q_{rad}$, in $\text{W}$), as schematized in Fig. 1(b). For example, an extreme case in which the membrane would be perfectly coupled to its environment via radiation would yield $q_{cond}=0$, and $q_{rad}=q$, where $q=\dot qV$ and $V$ is the membrane volume. We neglect the contribution of convective heat transfer, an assumption representative of a system in vacuum as in most optomechanics and thermal radiation sensing experiments. In steady state, heat flux out of the system must equal heat generation inside the system, yielding the general heat equation:

\begin{equation}
-k\cdot \nabla^2 T_{m}+\dot q_{rad} = \dot q,
\label{eq:one}
\end{equation}
where $k$ is the membrane conductivity, and $T_{m}$ is the position-dependent membrane temperature. In  Eq.~(\ref{eq:one}), $\dot q_{rad}$ is the radiative exchange, per unit volume, between membrane and the environment at $T_{\infty}$; is it given by:

\begin{equation}
\dot q_{rad}=\frac{2\sigma\epsilon(T_{m}^4-T_{\infty}^4)}{d},
\label{eq:two}
\end{equation}

Where $d$ is the membrane thickness, $\sigma$ is Stefan-Boltzmann constant, and $\epsilon$ is the total hemispherical emissivity of SiN. The factor 2 accounts for emission on both faces of the membrane. We consider Dirichlet boundary conditions, setting the temperature of the SiN membrane edges to be the same as the ambient temperature ($T_{\infty}$).

The non-linear nature of radiative heat exchange ($\dot q_{rad}\propto T_m^4$) prevents direct derivation of a closed-form solution of the heat equation. We consequently linearize Eq.~(\ref{eq:two})  by considering a small temperature difference between the membrane and the environment, yielding:
\begin{equation}
\dot q_{rad} \approx \frac{8\sigma\epsilon T_{\infty}^3}{d}(T_m-T_{\infty}).
\label{eq:three}
\end{equation}
Due to the minute scale thickness of the SiN film, we consider uniform temperature along the direction normal to the surface, thus reducing Eq.~(\ref{eq:one}) to a two-dimensional problem. 

For a circular membrane of radius $r_0$, the solution to the linearized heat equation in cylindrical coordinates is conveniently simple:

\begin{equation}
T_{m}(r) = [1-\frac{I_0(\beta \cdot r)}{I_0(\beta \cdot r_0)}]\cdot \frac{\dot q}{k\cdot \beta^2}+T_{\infty},
\label{eq:four}
\end{equation}
where
\begin{equation}
\beta = \sqrt{\frac{8\sigma\epsilon T_{\infty}^3}{kd}},
\label{eq:five}
\end{equation}
and $I_N$ is the $N^{th}$ order modified Bessel function of the first kind. From this temperature profile, we calculate heat transfer by conduction at the boundaries, using Fourier law of conduction:
\begin{equation}
q_{cond}=-2k\pi r_0 d\cdot\frac{\partial T_m}{\partial r}\Bigr|_{r_0},
\label{eq:six}
\end{equation}
which yields:
\begin{equation}
q_{cond}=\frac{2q}{\beta \cdot r_0}\cdot\frac{I_1(\beta \cdot r_0)}{I_0(\beta \cdot r_0)}.
\label{eq:seven}
\end{equation}
From Eq.~(\ref{eq:seven}), we can finally calculate the fraction of heat that leaves the membrane radiation ($x_{rad}$):
\begin{equation}
x_{rad}=\frac{q_{rad}}{q} = 1-\frac{q_{cond}}{q}=1-\frac{2}{\beta \cdot r_0}\cdot\frac{I_1(\beta \cdot r_0)}{I_0(\beta \cdot r_0)}.
\label{eq:eight}
\end{equation}\
In Eq.~(\ref{eq:eight}), we note that the right-hand side depends only on intrinsic membrane properties (i.e., independent of $\dot q$). As such, solving for $x_{rad}>0.5$ yields the properties required for a membrane to be thermally coupled to its environment more strongly via radiation than via solid-state conduction.

From the temperature profile of the membrane [see Eq.~(\ref{eq:four})], we can also express the thermal time constant of the membrane ($\tau$, in s) and its overall thermal conductance with the environment ($G$, $\textrm{in} \mathrm{W/K}$), both of which are of particular importance for use in sensors \cite{Rogalski2003}, and for predicting noise profiles in micro resonators \cite{Sansa2016,Fong2012,Cleland2002}. Note that $G = G_{cond}+G_{rad}$ includes heat transfer both by conduction in the supporting frame ($G_{cond}$) and by radiation ($G_{rad}$). We obtain $G$ and $\tau$ by evaluating the average temperature ($\overline{T}_m$) of the membrane [i.e., by integrating Eq.~(\ref{eq:four})] and using: 

\begin{eqnarray}
G = \frac{q}{\overline{T}_m-T_{\infty}}=\frac{k\pi(\beta\cdot r_0)^2 d}{1-\frac{2}{\beta \cdot r_0}\cdot\frac{I_1(\beta \cdot r_0)}{I_0(\beta \cdot r_0)}},
\label{eq:nine}\\
\tau = \frac{c_p\rho V}{G}=\frac{c_p \rho[1-\frac{2}{\beta \cdot r_0}\cdot\frac{I_1(\beta \cdot r_0)}{I_0(\beta \cdot r_0)}]}{k\beta^2},
\label{eq:ten}
\end{eqnarray}
where $c_p$ and $\rho$ are respectively the specific heat capacity and the material density. We note that, when heat transfer becomes entirely dominated by radiation (i.e. for large areas), Bessel terms become negligible and both expressions simplify to radius-independent quantities: 

\begin{eqnarray}
G_{rad}=8A\sigma\epsilon T_{\infty}^3,
\label{eq:eleven}\\
\tau_{rad}=\frac{\rho c_p d}{8\sigma\epsilon T_{\infty}^3}.
\label{eq:twelve}
\end{eqnarray}

For a square membrane, a geometry much more frequently encountered in practice, an analytical solution to the linearized heat equation also exists, but comprises an infinite amount of Fourier terms in order to respect the boundary conditions ($T=T_{\infty}$ at the membrane edges). Rather than using this complex solution, we numerically solve for the temperature profile of a square membrane using finite element analysis. From this solution (see supplemental Fig. S1), we determine that the radiative thermal coupling ratio ($x_{rad}=q_{rad}/q$) of a square membrane of side length $L$ matches that of a circular membrane if we consider an effective radius:
\begin{equation}
r_{eff}=1.252\frac{L}{2}.
\label{eq:thirteen}
\end{equation}
Logically, this effective radius falls between half of a square membrane side length ($L/2$), and half of its diagonal ($L/\sqrt{2}$), as shown schematically in Fig. 1(a). By replacing Eq.~(\ref{eq:thirteen}) in Eq.~(\ref{eq:eight}), the fraction of heat transfer occurring by radiation in a square membrane ($x_{rad}$) is given by:
\begin{equation}
x_{rad}=1-\frac{2}{\beta \cdot r_{eff}}\cdot\frac{I_1(\beta \cdot r_{eff})}{I_0(\beta \cdot r_{eff})}.
\label{eq:fourteen}
\end{equation}

The total thermal conductance of a square membrane is subsequently:
\begin{equation}
G = \frac{G_{rad}}{x_{rad}}=\frac{8L^2\sigma\epsilon T_{\infty}^3}{x_{rad}},
\label{eq:fifteen}
\end{equation}
while $\tau$ is now given by:
\begin{equation}
\tau = \frac{c_p \rho d}{8\sigma\epsilon T_{\infty}^3} x_{rad}.
\label{eq:sixteen}
\end{equation}
Eq.~(\ref{eq:sixteen}) yields an aberrant $\tau \approx 0$ when $x_{rad}\approx0$, but this result occurs for $L\approx0$, in which case our assumption of a 2-D problem (i.e., $L>>d$) does not hold. 

\section{Emissivity of SiN films}

From Eq.~(\ref{eq:eight}) to Eq.~(\ref{eq:sixteen}), it is obvious that the total hemispherical emissivity ($\epsilon$) is a key parameter governing radiative heat transfer in free-standing membranes. We calculate $\epsilon$ for the specific case of SiN using Kirchhoff law--i.e., directional spectral emissivity is equal to absorption:
\begin{equation}
\epsilon_{\lambda,\theta}(\lambda,\theta)=\alpha_{\lambda,\theta}(\lambda,\theta).
\label{eq:seventeen}
\end{equation}
We calculate $\alpha_{\lambda,\theta}(\lambda,\theta)=1-R_{\lambda,\theta}(\lambda,\theta)-T_{\lambda,\theta}(\lambda,\theta)$ using conventional optical multi-layer calculation \cite{emissivity} with the complex permittivity of SiN taken from \cite{Cataldo2012}. $R$ and $T$ denote the optical power reflection and transmission coefficients, respectively. We find that these coefficients depend non-negligibly on the angle, such that we integrate the emissivity according to Lambert's cosine law to obtain \cite{L.Bergman2017}:
\begin{equation}
\epsilon_{\lambda}(\lambda)=2\int_{0}^{\pi/2} \epsilon_{\lambda,\theta}(\lambda,\theta)\cdot cos(\theta)\cdot sin(\theta) d\theta,
\label{eq:eighteen}
\end{equation}

\begin{figure}[!htb]
    \centering
    \includegraphics[width=0.49\textwidth]{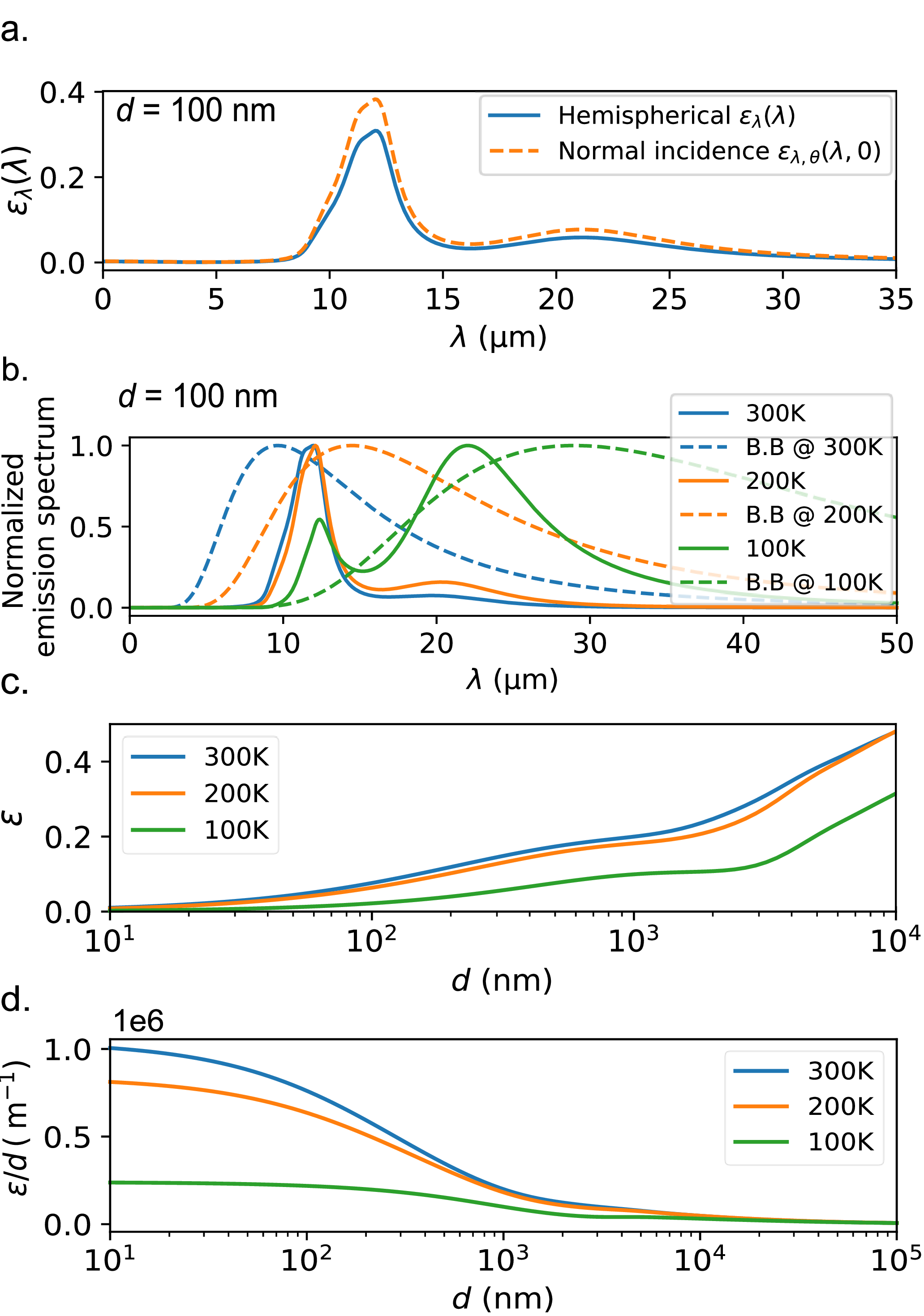}
    \caption{\label{fig:two} (a) Comparison between normal and hemispherical emissivity for free-standing SiN membranes. (b) Normalized emission spectrum of a 100 nm thick SiN membrane at different temperatures. Dashed lines represent blackbody spectra. As temperature decreases, the peak of the emission spectrum moves towards longer wavelengths. (c) Hemispherical total emissivity of free-standing SiN membranes at different temperatures. (d) Hemispherical total emissivity normalized by film thicknesses for different temperatures. This quantity scales with of the fraction of heat transfer occurring by radiation ($x_{rad}$) and is maximized for lower membrane thicknesses.}
 
\end{figure}
which is plotted in Fig. 2(a) for the specific example of a 100 nm thick membrane. In comparison, the normal directional spectral emissivity $\epsilon_{\lambda,\theta}(\lambda,0)$ is roughly $25\%$ higher than the integrated value $\epsilon_\lambda(\lambda)$ . In other words, emission and absorption are slightly stronger at normal incidence. This may be beneficial for radiation sensing applications \cite{Rogalski2003} where a sensor should be more strongly coupled at normal incidence (i.e. pointing at the object to be detected). We finally obtain the total hemispherical emissivity $\epsilon$ by weighting $\epsilon_\lambda(\lambda)$  with the blackbody emission spectrum $E_{\lambda,b}(\lambda,T)$ at temperature $T$:
\begin{equation}
\epsilon(T)=\frac{\int_{0}^{\infty}\epsilon_\lambda(\lambda)\cdot E_{\lambda,b}(\lambda,T)d\lambda}{E_b(T)},
\label{eq:nineteen}
\end{equation}
where $E_b(T)=\sigma T^4$. For concision, we use the notation $\epsilon(T)=\epsilon$ in this work. This weighting is shown, for various membrane temperatures, in Fig. 2(b) while $\epsilon$ as a function of thicknesses, is presented in Fig. 2(c). We find that both the emission distribution [Fig. 1(b)] and $\epsilon$ [Fig. 1(c)] weakly depend on temperature unless cryogenic membrane temperatures (e.g., 100 K) are considered.We willingly omit calculations at very low temperatures ($< 1 \,\mathrm{K}$) as this would presumably require a different model for material properties of SiN. We also note that the hemispherical emissivity for thin films is a strong function of thickness. This emphasizes that the common assumption of $\epsilon\approx0.6$ \cite{Zhang2013} for SiN appears appropriate only for bulk SiN materials and not for thin films.

We note that thicker films lead to higher emissivity [see Fig. 2(c)]--however as shown in Eq.~(\ref{eq:five}), the relevant quantity for estimating the contribution of radiation to the total heat transfer is the ratio $\epsilon/d$. This is plotted in Fig. 2(d), from which we conclude that, for a given surface area, thinner membranes are more efficiently coupled via radiation. The thickness dependence is the strongest in the 100 nm--1 $\mathrm{\mu m}$  range, while the relation reaches a relative plateau for thickness commonly employed in opto-mechanics experiments ($d<100\,\mathrm{nm}$) \cite{Zwickl2008}.
\begin{figure}[!htb]
    \centering
    \includegraphics[width=0.50\textwidth]{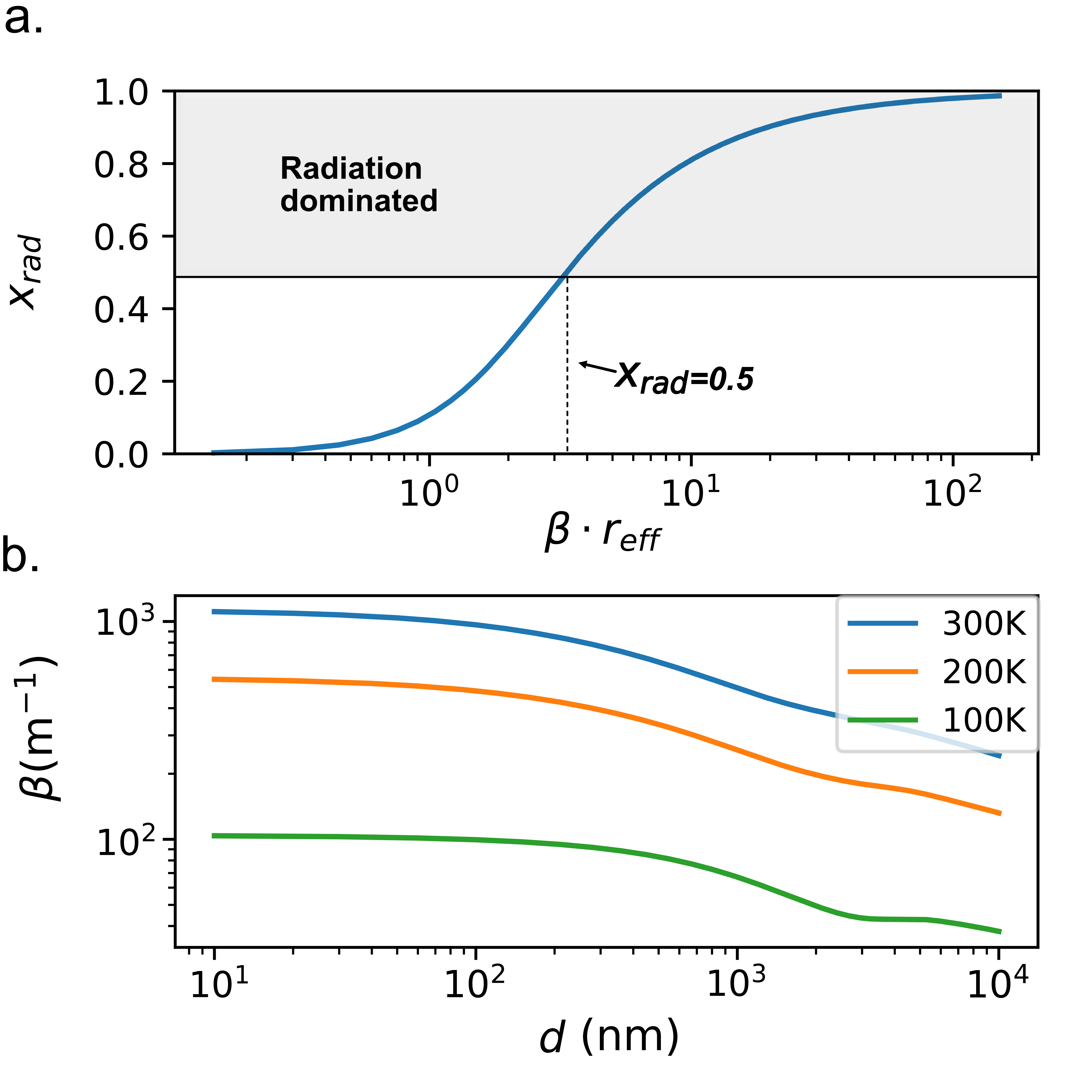}
    \caption{\label{fig:three} (a) Fraction of total heat transfer occurring via radiation ($x_{rad}$) in free-standing membranes as a function of $\beta\cdot r_{eff}$. The model applies both to circular ($r_{eff}=r_0$) and square ($r_{eff}=1.252 L/2$) membranes. For $x_{rad}$ larger than 0.5, membrane thermal coupling is radiation-dominated. (b) $\beta$ as a function of membrane thickness for the specific case of SiN}
 
\end{figure}

Having calculated the $\epsilon/d$ ratio, we can determine the critical membrane length for which heat transfer is dominated by radiation (i.e., $x_{rad}>0.5$). Using the closed-form relation developed in Eq.~(\ref{eq:fourteen}), we plot $x_{rad}$ as a function of $\beta\cdot r_{eff}$ in Fig. 3(a). The $\beta$ value for SiN, for given $d$ and $T_m$, is given in Fig. 3(b). We note that Fig. 3(a) is a universal relation for any freestanding thin film membrane anchored on all sides, while Fig. 3(b) accounts for the specific case of SiN. In Fig. 3(a), we find that thermal coupling of free-standing thin films is radiation dominated for $\beta\cdot r_{eff}\approx3.33$. From this, we finally derive a simple expression for the threshold of a radiation-coupled membrane:

\begin{equation}
r_{eff} > \frac{3.33}{\beta}=3.33\sqrt{\frac{kd}{8\sigma\epsilon T_{\infty}^3}},
\label{eq:twenty}
\end{equation}
for which the $\epsilon/d$ ratio can be obtained graphically from Fig. 2(d), in the specific case of SiN. We note that this equation is the same for a circular membrane, using $r_{eff}=r_0$ in lieu of $r_{eff
}=1.252 L/2$ for a square membrane.

\section{Experimental results}
We verify our model by conducting experiments using a commercially available ($3\times3$ mm side length, 200 nm thickness) low stress SiN membrane. We correlate the membrane temperature to its mechanical resonance frequency ($f_r$) using the relation given in \cite{St-Gelais2019}. The change in temperature of the membrane ($\Delta T$) is measured by comparing its instantaneous resonance ($f_r$) with its initial room-temperature resonance ($f_0$):
\begin{equation}
\frac{f_r}{f_0}=\sqrt{1-\frac{E\alpha}{\sigma_0}\Delta T}
\label{eq:twentyone}
\end{equation}
where SiN Young's modulus is $E=300\, \textrm{GPa}$, the thermal expansion coefficient is $\alpha=3.27\times10^{-6}\,\mathrm{K^{-1}}$ \cite{Tien2012} and the tensile stress is $\sigma_0\approx100\,\mathrm{Mpa}$. We consider a $25\%$ uncertainty for this relation given the variability of material constants and on the membrane dimensions [see error bars in Fig. 4(d)]. The experiment is conducted in high vacuum ($1.5\times10^{-6}\,\mathrm{torr}$) to eliminate convection heat transfer and viscous damping by air.

\begin{figure}[!htb]
    \centering
    \includegraphics[width=0.46\textwidth]{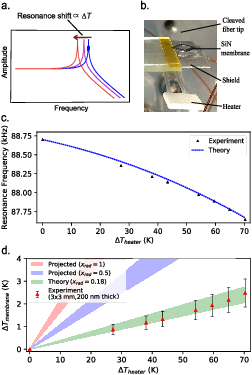}
    \caption{\label{fig:four} (a) Schematic of temperature-dependent frequency shift mechanism for SiN membranes. (b) Photograph of the experimental setup inside the vacuum chamber. For the actual experiment, the heater is placed closer (5 mm) from the membrane but is displaced here for a clearer picture (c) Resonance frequency of mechanical mode 2, 2 as a function of heater temperature. (d) Membrane temperature as a function of heater temperature. Experimental results agree with our model within a 25\% error on the membrane material constants. Predicted theoretical results for $x_{rad}=0.5$ and 1 are also shown for illustrative purpose, for the same $F = 0.55$ view factor. Shaded areas are bounded by the range $\epsilon_{heater}=0.6$ and 0.8.}

\end{figure}

We vary the temperature of the SiN membrane by exposing it to a rough-surfaced aluminum oxide heater [see Fig 4.(b)] placed within a short distance (5 mm) from the membrane. We infer the heater temperature by correlating it to its electrical resistance using a temperature coefficient of resistance of $4.7\times10^{-3}\,\mathrm{K^{-1}}$. We measure this value in a separate experiment by placing the heater on a hot plate and by measuring its resistance as a function of temperature. Due to the variability in documented value for the total hemispherical emissivity of rough-surfaced aluminum oxide \cite{L.Bergman2017}, we assume an upper bound value of $\epsilon_{heater}=0.8$ and a lower bound of 0.6. A reflective aluminum shield is placed between the membrane and the heater to prevent the silicon frame of the membrane from heating up upon absorption of radiation. A piezoelectric actuator is attached to the same glass slide as the membrane to excite its mechanical resonance. Membrane displacement is measured using an optical interferometer setup \cite{Rugar1989}. The instantaneous resonance ($f_r$) of the SiN membrane shifts by over 1 KHz when subject to a 70 K increment in heater temperature, as presented in Fig. 4(c).

By neglecting photons having more than one interaction with the heater and membrane (due to the relatively small size of the membrane, the diffusive surface of the heater, and its high emissivity/absorptivity) the temperature of the membrane is correlated to the heater temperature simply by:
\begin{equation}
\Delta T_m \approx \frac{\Delta T_{heater}}{2}\cdot x_{rad}\cdot F\cdot\epsilon_{heater}
\label{eq:twentytwo}
\end{equation}
where, $F=0.55$ is the geometrical view factor \cite{L.Bergman2017}. The detailed thermal equivalent circuit from which Eq.~(\ref{eq:twentytwo}) derived is given in supplementary Fig. S2. From Eq.~(\ref{eq:twentytwo}), correlating the membrane temperature with the heater temperature allows measurement of $x_{rad}$ and validation of our model. This is presented in Fig. 4(d), where the experimental points fall within the theoretically expected values from Eq.~(\ref{eq:fourteen}). For illustrative purpose, we also plot the projected values for a membrane that would be on the threshold ($x_{rad}=0.5$) or completely radiation-dominated ($x_{rad}=1$), for the same geometrical view factor ($F=0.55$).  

\section{conclusion}
We expect that our work will be interest for achieving high performance radiation sensors exploiting the high temperature sensitivity of SiN mechanical resonance \cite{St-Gelais2019}. In such sensors, one would ideally want to achieve $x_{rad}\approx1$ to reach the highest possible detectivity \cite{Rogalski2003}. Our work shows that achieving such $x_{rad}$ value is feasible using realistic membrane dimensions. We readily achieve $x_{rad}=0.18$ using commercially available SiN membranes of non-optimized dimensions. We also expect that the provided closed-form expressions for $\tau$ and $G$ will be of great use for predicting frequency noise in high Q-factor SiN \cite{Sansa2016,Fong2012,Cleland2002}. Given the high temperature sensitivity of resonance frequency in SiN resonators, we expect temperature fluctuation noise--which is directly linked to $\tau$ and $G$ \cite{Cleland2002}--to have a non-negligible contribution to frequency noise. Finally, we expect that outlining the non-negligible contribution of radiation heat transfer may be useful in experiments involving cooling of SiN membranes. As a striking example, a membrane with $x_{rad}>0.5$ could be more efficiently cooled by a cold object facing it, than by direct contact cooling of its supporting silicon frame. 
\bibliography{ISARC}
\clearpage

\section*{Supplementary Information}
\addcontentsline{toc}{section}{Supplementary Information}
\renewcommand{\thesection}{S}
\renewcommand{\theequation}{S\arabic{equation}}
\renewcommand{\thefigure}{S\arabic{figure}}
\subsection{Numerical simulation}

The 2-D steady-state temperature profile of the squre-sized SiN membrane is simulated using MATLAB PDE solver, such that the closed-form analytical results can be compared with the numerical simulation. We first define the computational geometry to be a square which has the same side length (3 mm) as the sample SiN membrane. We then specify the coefficients for the PDE model by rearranging the heat equation as:

\begin{figure}[!htb]
    \setcounter{figure}{0}
    \centering
    \includegraphics[width=0.46\textwidth]{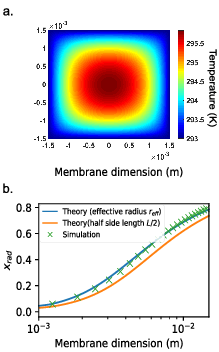}
    \caption{\label{fig:four} (a) 2-D temperature simulation for $3\times3$mm, 200 nm SiN membrane. (b) $x_{rad}$ as a function of side length which discrete points are given by the numerical simulation, whereas the continuous lines are plotted by closed-form expression.}
\end{figure}
\begin{equation}
-\nabla\cdot(\nabla T_m)+{\beta^2} T_m = \frac{\dot q}{k}+\beta^2 T_{\infty}
\setcounter{equation}{1}
\label{eq:twentyone}
\end{equation}
where $\sigma$ is the Stefan-Boltzmann constant, $\epsilon$ is the total hemisphereical emissivity of the 200 nm ($d$) SiN membrane which is calculate to be 0.11, $T_{\infty}$ is the ambient temperature (293 K), $k$ is the conductivity of the SiN membane which is found to be 12 $\mathrm{W}/\mathrm{m}\cdot\mathrm{K}$. Note that $\dot q$ is the heating power being absorbed by the SiN membrane per unit volume. Here we arbitrarily set $\dot q = 6.5 \times 10^7 \mathrm{W/m^3}$. The simulated temperature profile is shown in Fig. S1(a).

With the numerical simulation, we can obtain the fraction of heat that leaves the membrane by radiation ($x_{rad}$) using the simulated temperature profile. By plotting this result and the analytical results for many side lengths values, we find that the correction ratio between the effective radius and half of the side length to be 1.252 which is presented in Eq.~(\ref{eq:thirteen})
\subsection{Thermal equivalent circuit}
\begin{figure}[!htb]
    \setcounter{figure}{1}
    \centering
    \includegraphics[width=0.46\textwidth]{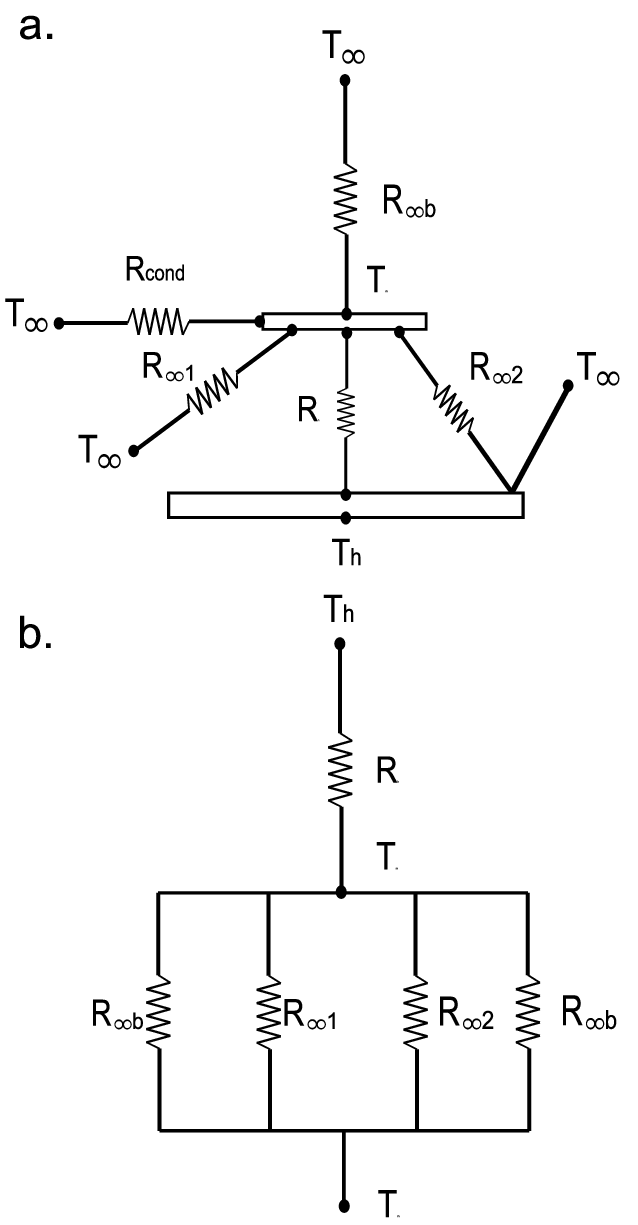}
    \caption{\label{fig:four} Thermal equivalent circuit considering the heater, the membrane and the vacuum environment. (a) Thermal circuit schematic. (b) Simplified thermal circuit.}
\end{figure}
The analogy of the electric circuit is useful to understand the heat transfer of the SiN membrane with its environment. Here we neglect multiple heat exchanges between the SiN membrane and the heater by only considering single photon interaction. In other words, photons leaving the membrane and reflected by the surface of the heater will be directed to the environment, which results in resistance $R_{\infty 2}$ in Fig. S2. This assumption is based on the fact that the surface of the heater is highly diffusive, and the surface area of the membrane is substantially smaller than the area of the heater. Hence, photons emitted by the membrane and then hit the surface of the heater have minimal chance of being reflected back to the membrane.

By rearranging the thermal circuit schematic in Fig. S2(a), we reach the simplified thermal circuit shown in Fig. S2(b) for which the resistances are expressed as:
\begin{equation}
R_h = \frac{1}{\theta_{rad}\epsilon_{m}\epsilon_{h}F},
\setcounter{equation}{2}
\label{eq:twentyone}
\end{equation}
\begin{equation}
R_{\infty,b} = \frac{1}{\theta_{rad}\epsilon_{m}},
\setcounter{equation}{3}
\label{eq:twentyone}
\end{equation}
\begin{equation}
R_{\infty,1} = \frac{1}{\theta_{rad}\epsilon_{m}(1-F)},
\setcounter{equation}{4}
\label{eq:twentyone}
\end{equation}
\begin{equation}
R_{\infty,2} = \frac{1}{\theta_{rad}\epsilon_{m}F(1-\epsilon_h)},
\setcounter{equation}{5}
\label{eq:twentyone}
\end{equation}
\begin{equation}
\theta_{rad}=4\sigma A_m T_{\infty}^3.
\setcounter{equation}{6}
\label{eq:twentyone}
\end{equation}

Here, $\epsilon_m$ is the total hemispherical emissivity of the SiN membrane, $\epsilon_h$ is the total hemispherical emissivity of the ceramic heater, $F$ is the geometrical view factor from the membrane to the heater and $A_m$ is surface area of the SiN membrane. By following the thermal circuit in Fig. S2(b), we obtain Eq.~(\ref{eq:twentytwo}) in the main text.

\end{document}